\documentclass[sn-nature,Numbered,iicol]{sn-jnl}% Basic Springer Nature Reference Style/Chemistry Reference Style

\usepackage{graphicx}%
\usepackage{multirow}%
\usepackage{amsmath,amssymb,amsfonts}%
\usepackage{amsthm}%
\usepackage{mathrsfs}%
\usepackage[title]{appendix}%
\usepackage{xcolor}%
\usepackage{textcomp}%
\usepackage{manyfoot}%
\usepackage{booktabs}%
\usepackage{algorithm}%
\usepackage{algorithmicx}%
\usepackage{algpseudocode}%
\usepackage{listings}%

\begin{document}
%[b]%Blank page needed to appear here to push title to the right hand page[/b]

%___________________________________________________________________________
%%% Für Einreichen bei LSA: Daten der Autoren 

% \section*{hier Titel einfügen }

% Angaben zu Autoren 
% \pagenumbering{Roman}% Capital 'R': uppercase Roman numerals
% \newpage

%%%
%___________________________________________________________________________

\title[Continuous adiabatic FMCW-LiDAR]{\textbf{Continuous adiabatic frequency conversion for FMCW-LiDAR}}


% \author*[1]{\fnm{Alexander} \sur{Mrokon}} \email{alexander.mrokon@imtek.uni-freiburg.de, +49 761 203-98661}
% \author[1]{\fnm{Johanna} \sur{Oehler}}\email{johanna.oehler@posteo.de}
% \author[1,2]{\fnm{Ingo} \sur{Breunig}}\email{ingo.breunig@imtek.uni-freiburg.de}

\author*[1]{\fnm{Alexander} \sur{Mrokon}} \email{alexander.mrokon@imtek.uni-freiburg.de}
\author[1]{\fnm{Johanna} \sur{Oehler}}%\email{\nomail}
\author[1,2]{\fnm{Ingo} \sur{Breunig}}%\email{\nomail}

\affil[1]{\orgdiv{Laboratory for Optical Systems, Department of Microsystems Engineering - IMTEK}, \orgname{University of Freiburg}, \orgaddress{\street{Georges-Köhler-Allee 102}, \city{Freiburg}, \postcode{79110}, \country{Germany}}}

\affil[2]{\orgname{Fraunhofer Institute for Physical Measurement Techniques IPM}, \orgaddress{\street{Georges-Köhler-Allee 301}, \city{Freiburg}, \postcode{79110}, \country{Germany}}}


\abstract{
Continuous tuning of the frequency of laser light is the cornerstone for a plethora of applications in basic science as well as in the industrial environment. They range from the hunt for gravitational waves, the realization of optical clocks over health and environmental monitoring to distance measurements. So far, it is difficult to combine a wide tuning range ($>100$~GHz) with sub-microsecond tuning times, intrinsic tuning linearity and coherence lengths beyond 10~m. We show that electro-optically driven adiabatic frequency converters using high $Q$ microresonators made of lithium niobate are able to transfer arbitrary voltage signals into frequency chirps on time scales far below 1~\textmu s. The temporal variation of the frequency nicely agrees with the one of the voltage applied. Superimposing the converted light with the unconverted one, we derived from the beat signal that 200-ns-long linear frequency chirps deviate less than 1~\% from perfectness without any additional measures. The Coefficient of Determination is $R^2>0.999$. The coherence length of the emitted light is of the order of 100~m. In order to prove these findings, we apply linear frequency sweeps for FMCW LiDAR for distances between 0.5 and 10~m without any signs of nonlinearity. Combined with the demonstrated ns tuning, with the potential to tune the eigenfrequency of lithium-niobate-based resonators by several 100~GHz, our results show that electro-optically driven adiabatic frequency converters can be used in applications that require ultrafast and flexible continuous frequency tuning with intrinsic linearity and large coherence length.
}

%%================================%%
%% Sample for structured abstract %%
%%================================%%

% \textbf{Conclusion:} The abstract serves both as a general introduction to the topic and as a brief, non-technical summary of the main results and their implications. The abstract must not include subheadings (unless expressly permitted in the journal's Instructions to Authors), equations or citations. As a guide the abstract should not exceed 200 words. Most journals do not set a hard limit however authors are advised to check the author instructions for the journal they are submitting to.}




\maketitle
\pagenumbering{arabic} % Arabic/Indic page numbers
% \section{Main novelty Claim}\label{sec0}
% \textbf{\noindent\textcolor{red}{Main novelty Claim: AFC with high frequency agility and linear mode hop free tuning with first ever demonstration of an application in LiDAR}\\}

\section{Introduction}\label{sec1}

Since more than half a century, scientists work on controlling the emission frequency of lasers. In particular, continuous (mode-hop free) tuning is of prime interest for many applications. It enables the active stabilization of laser light to a target frequency value as well as to sweep it. Both are crucial for various applications. The detection of gravitational waves relies on an interferometer driven by a frequency-stabilized laser \cite{WilkeStabilized}. Exoplanets are found by applying stabilized frequency combs \cite{VahalaSearching}. Optical clocks can be realized by cooling and trapping atoms with frequency-stabilized laser light \cite{HidetschiAnOptical}. Health and environmental monitoring is based on high-resolution laser spectroscopy of naturally occurring molecules \cite{PanAdvances}. Artificial (photonic) molecules can be investigated as well \cite{KulakovskiiOpticalModes}. Distances and velocities are determined using frequency modulated continuous-wave (FMCW) LiDAR \cite{BoserLidar}. Furthermore, swept source optical coherence tomography (OCT) is heavily used for medical imaging \cite{FujimotoTheEcosystem}. Thus, the control of the frequency of laser light is not just crucial for basic science but as well as for out-of-the-lab applications.
 
%Frequency-tunable lasers have arisen as essential instruments, finding widespread utility across various areas of modern technological applications.
%In telecommunications, they enable high-capacity data transmission through wavelength-division multiplexing (WDM), supporting modern internet infrastructure. 
%Scientific research in many disciplines benefits from tunable lasers in advanced spectroscopy, aiding environmental monitoring, pharmaceutical development, and materials analysis \cite{PanAdvances}. 
%Healthcare relies on them for medical imaging techniques, with around 30 Million optical coherence tomography (OCT) imaging procedures performed worldwide every year, enhancing early disease detection and patient care \cite{FujimotoTheEcosystem}. 
%Their adaptability and precision are instrumental in harnessing the principles of quantum mechanics in the second quantum revolution for technological advancements such as quantum computing and high-precision quantum sensing \cite{MilburnQuantum}.

There are several approaches for continuous tuning of the frequency of laser light. The possible fields of application are determined by the following parameters: tuning range, tuning time, tuning linearity and coherence length. The ratio of the first two gives the tuning rate. In general, we can divide the tuning approaches into two basic concepts: The first actively influences the laser itself. The second one converts the frequency of light emitted by a laser without influencing the light source. Now, we compare different methods regarding the abovementioned parameters.

%% DFB

Commercially available Distributed Feedback (DFB) lasers can be tuned via changing the driving current. Here, one typically achieves several 10s of GHz tuning range in microseconds. Thus, we have tuning rates around 10~GHz/\textmu s. However, this simple tuning mechanism is not linear \cite{Li20}. The coherence length is of the order of 10~m.
%with thermal tuning capabilities extending into the THz range, as well as electrical tuning spanning multiple GHz. The tuning speed, contingent upon the specific design and mechanism employed, can reach the MHz regime.
%This results in a chirp rate, which is the scan range in a specific scan time, of 1 GHz/$\mu$s. \cite{FermannFrequency-Modulated} 
Combining 12 DFB elements, researchers achieved 5560 GHz tuning in 3000~\textmu s. By actively linearizing the frequency sweep, this approach was applied for determining distances of up to 6~m with better than 30~\textmu m resolution via FMCW LiDAR \cite{NehmetallahLarge-Volume}.

%%VCSEL
Furthermore, commercially available micro-electro-mechanically tuned vertical-cavity surface emitting laser systems (MEMS-VCSELs) cover about 20~THz at a central wavelength at 1~\textmu m with in 2~\textmu s. % Thorlabs SL104071 
This results in $10^4$~GHz/\textmu s tuning rate. However, one has to untertake some effort to linearize the tuning, e.g. with reference interferometers or with programmed controllers \cite{MingHigh-Accuracy,JayaramanVCSEL,MingLaserFreq,IiyamaThreee-Dimensional}. The coherence length depends on the tuning range and on the tuning speed. Typically, it is in the 1-m range. However, restricting the tuning range or using low tuning rates, the coherence length can increase by two orders of magnitude \cite{Jayaraman14}. With proper tuning linearization, MEMS-VCSELS are applied for medical imaging via OCT or for long-distance measurement via FMCW LiDAR \cite{Khan21}.

%% Akin
Wide and linear frequency tuning under complete electronic control is achieved in semiconductor-based akinetic swept sources \cite{DrexlerAkinetic}. Here, five control voltages are synchronized in order to achieve a linear frequency sweep. These lasers provide 20~THz tuning range at a central wavelength around 1.3~\textmu m in a few microseconds. This leads to 4000 GHz/\textmu s tuning rate. The coherence lengths are of the order of 0.1~m. These lasers are applied for optical imaging via OCT \cite{DrexlerAkinetic,LeitgebAkinetic}.

%% RES
In order to combine large coherence length of hundreds of meters with a linear frequency tuning, one can couple a semiconductor laser or amplifier with an external high-quality-factor resonator. Tuning of the laser is done by sweeping the resonance frequency of the cavity applying the linear electro-optic effect (Pockels effect) \cite{KippenbergUltrafast,KippenbergHighDensity}. Using this approach, one achieves 2~GHz tuning range in 1~ns, i.e. 2000~GHz/\textmu s \cite{VahalaPockelsLaser}. Compared to the abovementioned methods, the major drawback is the limited tuning range. It is proportional to the quality factor of the tuned resonator \cite{Kondratiev17}. Thus, a high quality factor is beneficial for a large tuning range. However, it was demonstrated that the tuning time shall be considerably below the photon life time, i.e. in terms of tuning rate, the quality factor should be low \cite{VahalaPockelsLaser}. Thus, one cannot maximize tuning range and minimize tuning time at the same time. However, a successful application for FMCW LiDAR was demonstrated \cite{KippenbergUltrafast}.

To summarize, we see that none of the abovementioned approaches combines large tuning range, small tuning times, tuning linearity and high coherence length. Furthermore, they are all fundamentally limited to the gain bandwidth of the laser medium.

The second concept is based on conversion of the frequency of incident laser light. 
A promising approach is adiabatic frequency conversion (AFC). Here, laser light is coupled into an optical resonator. Then, the optical length of the cavity, i.e. its eigenfrequency, is varied on a time scale smaller than its decay time \cite{SatoshiWavelength}. The light stored in the resonator changes its frequency according to the change of the eigenfrequency. This process is the optical analogue to changing the pitch of sound by varying the length of a guitar string. In the first experimental demonstration, light at 1.5~\textmu m wavelength was coupled into a silicon ring resonator. A short laser pulse generates mobile charges in the silicon changing the refractive index of the material. This leads to 300-GHz frequency shifts in a picosecond, i.e. a tuning rate of $3\times10^{8}$ GHz/\textmu s \cite{LipsonChanging}. Thus, already in its first experimental demonstration, this scheme outperforms the other ones by several orders of magnitude in terms of tuning rate. 
However, in this configuration the adiabatic frequency change is proportional to the energy of the control pulse, a continuous linear frequency sweep would be difficult to realize. Furthermore, it restricts the frequency change to positive values and usage of a pulsed laser to drive the system is quite an effort.

In order to change this unsatisfactory situation and to unlock the potential of AFC for ultrafast, flexible and continuous frequency tuning of laser light, we have replaced laser-based charge injection by the linear electro-optic effect \cite{BreunigPockels}. This promises a fast and linear response as well as flexibility in the sign for the frequency change. Using this scheme and a millimeter-sized resonator made of lithium niobate, we have achieved 5~GHz tuning range in 5~ns, i.e. 1000~GHz/\textmu s tuning rate \cite{Breunig22}. Recently, electro-optically driven adiabatic frequency tuning of 14~GHz in 14~ps, i.e. $10^6$~GHz/\textmu s was demonstrated with a chip-integrated microresonator made of lithium niobate \cite{CardenasAdiabatic}. 

%Other implementations of AFC use diverse platforms such as photonic crystals, waveguides, and fiber grating cavities. \cite{FanDynamic,NotomiElectro,SatoshiWavelength,AgrawalEfficient,TanabeKerr}

%%______________________
\begin{figure*}[h]%
\centering
\includegraphics[width = \textwidth]{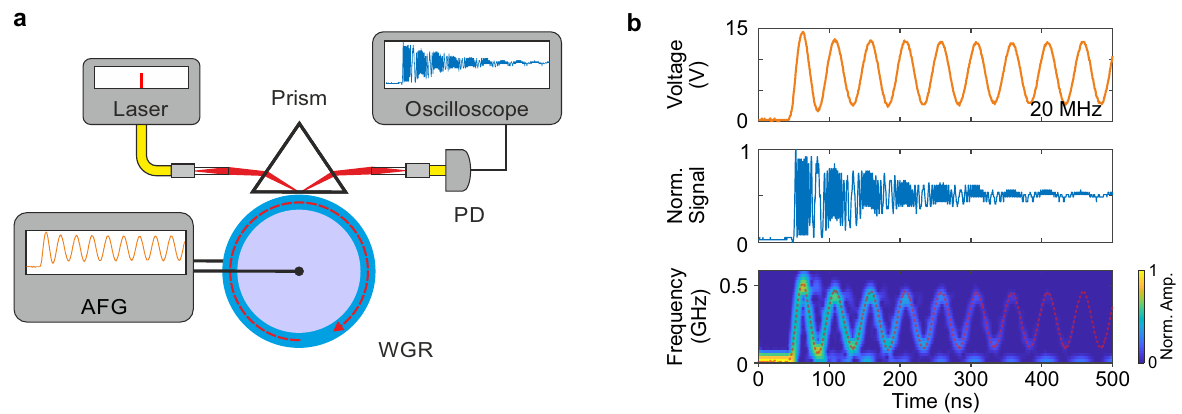}
\caption{\textbf{Electro-optically driven adiabatic frequency converter.} \textbf{a} Schematic setup including the pump laser, the coupling prism and the whispering gallery resonator (WGR). An arbitrary function generator (AFG) modulates the eigenfrequency of the cavity and consequently the frequency of the light coupled out of the prism. The beat signal between unconverted light and converted one is detected with a photodiode (PD) and displayed with an oscilloscope.
 \textbf{b} Temporal behaviors of a signal driving the AFG with 20~MHz, of the corresponding normalized beat signal and the derived frequency shifts. The dashed red line shows a linearly scaled version of the driving electrical signal.}\label{fig:Setup_Part1}
\end{figure*}
Since frequency tuning in this scheme is enabled by the linear electro-optic effect and the laser remains untouched, it should be possible to generate linear frequency sweeps with coherence lengths beyond 10~m. Thus, electro-optically driven adiabatic frequency conversion might combine ultrafast and flexible frequency tuning with high linearity and large coherence length. 
However, a more elaborate investigation of continuous adiabatic tuning is still missing. Furthermore it was not yet applied for any application that requires continuous tuning. Currently, one might even wonder whether adiabatic frequency converters are useful for any real-world application. In this work, we will investigate the continuous tuning of electro-optically driven adiabatic frequency converters. In particular, we focus on the linearity of the tuning mechanism. Furthermore, we apply this tuning scheme for FMCW LiDAR which sets high demands regarding the light source.

%%%%_________________________

\section{Results}\label{sec2}
\subsection{Tuning linearity}

An electro-optically driven adiabatic frequency converter shifts the frequency $\nu$ of laser light coupled into a microresonator with the thickness $d$ by \cite{BreunigPockels}
\begin{equation}
    \Delta\nu = \frac{1}{2}\nu n^2 r \eta \frac{U}{d}\;.\label{eq:FrequencyShift}
\end{equation}
In this expression, $n$ denotes the refractive index of the resonator material, $r$ its electro-optic coefficient and $U$ the voltage applied between the electrodes on the top and bottom faces of the resonator. With the factor $\eta$, we account for the deviation of the geometry from a plate capacitor \cite{MinetElectro-Optic}. Equation (\ref{eq:FrequencyShift}) indicates that a continuously varied voltage signal $U(t)$ is linearly transferred to a frequency shift $\Delta\nu(t)$ as long as the variation happens at a time scale smaller than the resonator's decay time. 

In order to verify this, we use the setup sketched in Fig.~\ref{fig:Setup_Part1}a. Laser light at $\nu=192$~THz (1560~nm wavelength) is coupled into a millimeter-sized whispering gallery resonator made of lithium niobate using a rutile prism. An arbitrary function generator provides the temporally varying voltage signal $U(t)$. The frequency-shifted light at $\nu+\Delta\nu(t)$ is coupled out of the resonator and interferes with the non-shifted laser light at $\nu$, forming a beat signal captured by a photodiode connected to an oscilloscope. From the beat signal, we extract the actual frequency change, i.e. $\Delta\nu(t)$ and compare its temporal behavior with $U(t)$. 

We start our analysis with a 500-ns-long signal varying between 0 and 15~V comprising a 20~MHz sinusoidal oscillation. Figure~\ref{fig:Setup_Part1}b shows this signal, the corresponding beat signal as well as the frequency shift. The beat signal clearly shows low beat frequencies at low voltages and high beat frequencies at high voltages. Furthermore, it reveals that the resonator has several 100~ns decay time. The temporal dependence of the frequency shift shows all features of the driving voltage signal. The 20-MHz oscillation is as obvious as the overshoot at the first oscillation, i.e. even tiny details of the voltage signal are transferred to the frequency shift. At 15~V driving signal, we observe 510~MHz frequency shift.
\begin{figure*}[h]%
\centering
\includegraphics[width = \textwidth]{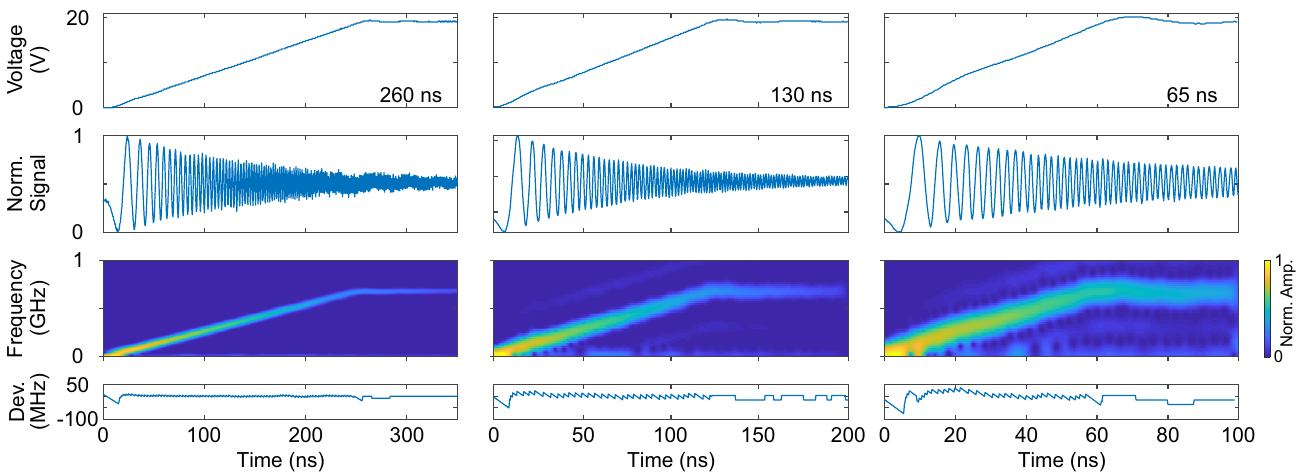}
\caption{\textbf{Generation of linear frequency chirps.} Driving voltages from the arbitrary function generator, corresponding normalized beat signals, the derived frequency changes and the deviation from a perfect linear frequency chirp. The respective signals are displayed for 260, 130 and 65 ns rise times.\label{fig:Chirps}}
\end{figure*}

For a more quantitative analysis, we aim for linear frequency chirps since they are of most importance for various applications. We set the arbitrary function generator to deliver driving signals in which the voltage increases linearly from 0 to 20~V in 260, 130 and 65~ns, respectively. 
They are displayed in Fig.~\ref{fig:Chirps} together with the corresponding normalized beat signals, the frequency changes as well as the deviations from perfectly linear frequency chirps. For 260 ns rise time, the function generator delivers an almost perfect linear voltage increase. The corresponding frequency change reveals a steady increase of the frequency from 0 to 690~MHz. The deviation from a perfectly linear chirp is lower than 10~MHz for times between 20 and 250~ns. Only from 0 to 20~MHz and from 250 to 260~ns, we observe a larger deviation. 
Across the entire time span, the rms value of the deviation is 5~MHz, while a linear fit results in a Coefficient of Determination (R-Squared),which is a measure of the goodness of fit, with  $R^2=0.999$.
The investigation of frequency chirps with smaller rise times (130 and 65~ns) and comparing them with the ones from the 260-ns-long chirp reveals the following:
The voltage signal delivered by the function generator increases less linearly from 0 to 20~V as the rise time decreases. The frequency increases continuously from 0 to 690~MHz in all cases. However, the uncertainty of the frequency change as well as the deviation from a perfectly linear chirp increases significantly as the rise time decreases. For 130 and 65~ns, we obtain rms values of 11 and 20~MHz as well as $R^2$ values of 0.995 and 0.985, respectively. Similar to the 260-ns-long chirp, the deviation of the instantaneous frequency seems to be significantly larger in the beginning of the chirp, i.e. at times smaller than 10~ns.

\subsection{Application for FMCW-LiDAR}
%\noindent\textcolor{blue}{Deviation berechnung schön und gut. Aber um zu zeigen, dass wirklich linear --$>$ Demonstration von Anforderung mit sehr hohen Ansprüchen an Linearität}\\
\begin{figure*}[h]%
\centering
\includegraphics[width = \textwidth]{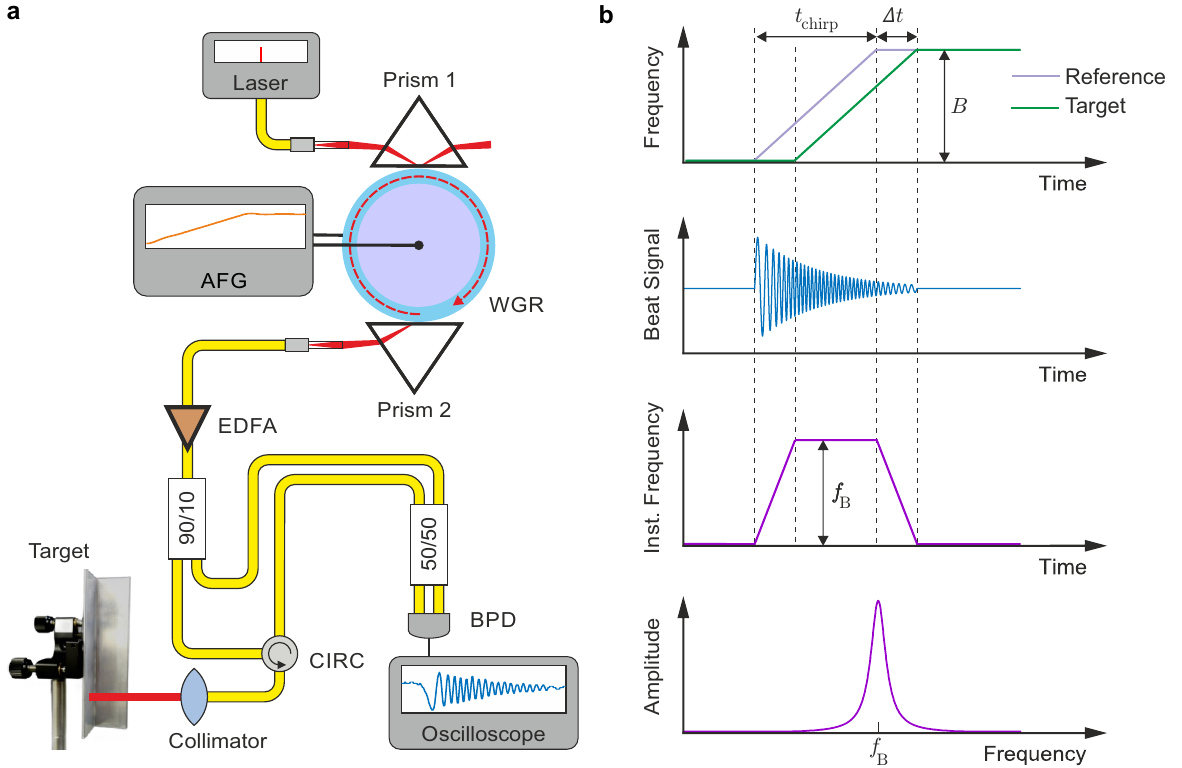}
\caption{\textbf{Adiabatic frequency converter for FMCW LiDAR.} \textbf{a} Experimental setup comprising an electro-optically driven adiabatic frequency converter sketched in Fig.~\ref{fig:Setup_Part1}. The frequency shifted light is coupled out using an additional prism (Prism 2). Optical fiber splitters and a circulator (CIRC) ensure that reference light and light reflected from the distant target form a beat signal at a balanced photodetector (BPD) connected to an oscilloscope. \textbf{b} Determination of the maximum beat frequency $f_\mathrm{B}$ which is proportional to the distance of the target. Interfering linearly chirped reference and target signals with the bandwidth $B$ form a beat signal with a temporally varying frequency having the maximum value $f_\mathrm{B}$. Fourier transform of the beat signal directly yields $f_\mathrm{B}$.}
\label{fig:Setup_Part2}
\end{figure*}
Our investigation of continuous adiabatic frequency tuning shows that we can generate frequency chirps with a high degree of linearity. Now, we want to adress the question, whether adiabatic frequency shifters can serve as useful light sources. For this purpose, we apply an electro-optically driven adiabatic frequency converter for determining distances of up to 10~m using FMCW LiDAR as a proof-of-concept. Here, the combination of linear frequency tuning and long coherence length are of paramount importance.

The experimental setup is sketched in Fig.~\ref{fig:Setup_Part2}a. It comprises the adiabatic frequency converter described above supplemented with a second prism. This ensures to couple out frequency-chirped light without letting it interfere with non-converted light. The chirp is characterized by the bandwidth $B$ and the chirp time $t_\mathrm{chirp}$, i.e. by the slope $B/t_\mathrm{chirp}$. About 10 percent of the output light serve as the reference signal whereas 90 percent of the output light are sent to a distant target, a non-polished metal plate. The light reflected from the target and the reference light are shifted in time by $\Delta t$ being proportional to the distance of the target. They form a beat signal detected by a balanced photodiode, visualized on an oscilloscope. The linearly chirped reference and target signals as well as the resulting beat signal are sketched in Fig.~\ref{fig:Setup_Part2}b. The frequency of the beat signal increases linearly from 0 to the maximum value $f_\mathrm{B}=(B/t_\mathrm{chirp})\Delta t$ during the time interval $\Delta t$, it remains constant at $f_\mathrm{B}$ for $t_\mathrm{chirp}-\Delta t$ before it decreases back to 0 during $\Delta t$ (see Fig.~\ref{fig:Setup_Part2}b). Thus, we have two options to determine $f_\mathrm{B}$ and consequently the distance from the target. We can analyze the temporal behavior of the instantaneous frequency or we can Fourier transform the beat signal. 

\begin{figure*}[h]%
\centering
\includegraphics[width = \textwidth]{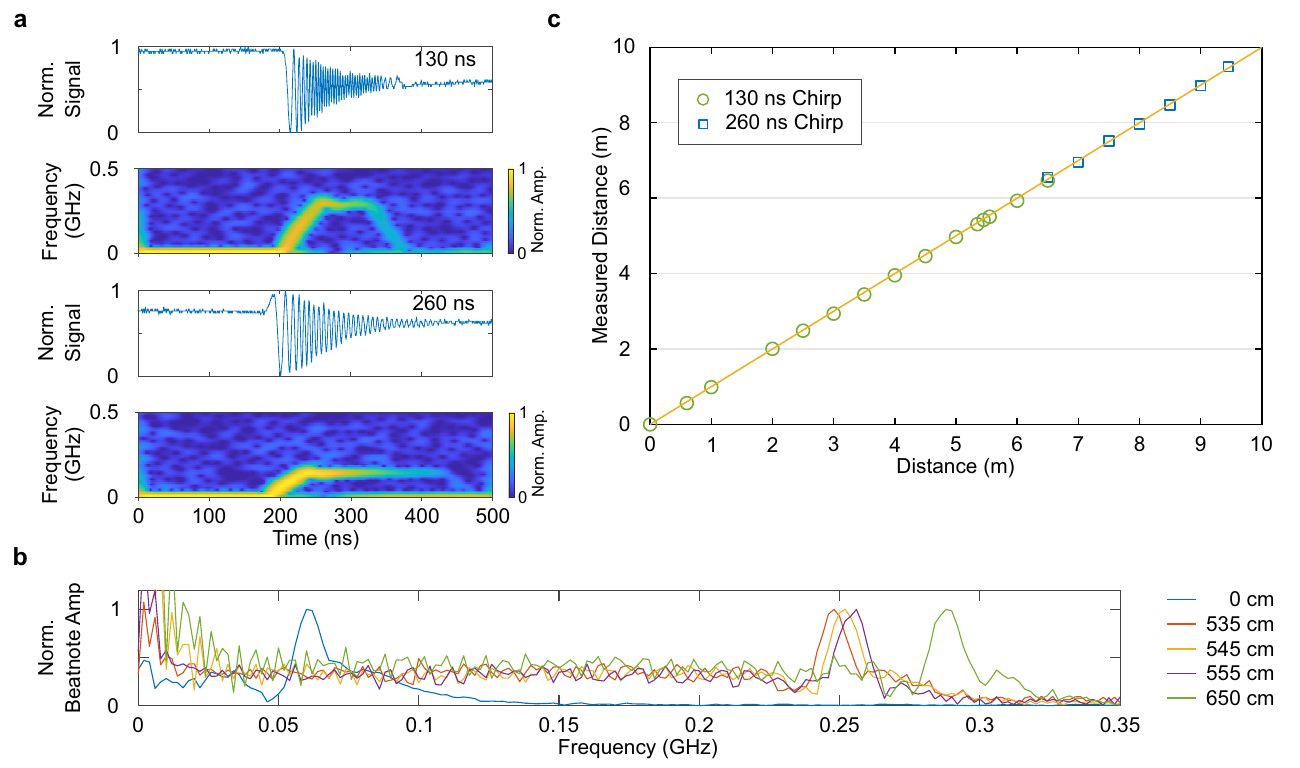}
\caption{\textbf{Distance measurement using frequency chirps from an adiabatic frequency converter.} \textbf{a} Beat signals and corresponding temporal evolution of the derived frequency shifts for 650~cm distance with 130 and 260 ns rise time, respectively. \textbf{b} Normalized Fourier transform of beat signals for different distances between 0 and 650~cm for 130~ns rise time. \textbf{c} Distance determined with AFC-based FMCW LiDAR (green circles and blue squares) vs. distance measured with a commercially available device based on intensity modulation. The yellow line represents a perfect correlation between the two measurement methods.  \label{fig:BeatSignal}}
\end{figure*}

Since the bandwidth of the balanced photodiode is limited to 300~MHz, we need to assure that the beat frequency does not exceed this value. Consequently, we perform the measurements at distances from 0 to 6.5~m with $t_\mathrm{chirp}=130$~ns and from 6.5 to 10~m with $t_\mathrm{chirp}=260$~ns. Figure~\ref{fig:BeatSignal}a shows the beat signals as well as the corresponding temporal evolution of the frequency shift for both rise times at 6.5~m distance. For both, the frequency shifts shows the temporal behavior already sketched in Fig.~\ref{fig:Setup_Part2}b. 
The only significant difference between the two is the magnitude of the maximum frequency. For 130~ns we observe 290~MHz and for 260~ns, the maximum frequency is halved. In Fig.~\ref{fig:BeatSignal}b we have displayed the Fourier transformation of beat signals for 130~ns rise time and distances between 0 and 6.5~m. For zero distance, we find a clear maximum at 60~MHz frequency and for 6.5~m one at 290~MHz. Evaluating the temporal behavior of the frequency shift and Fourier transform deliver the same values. The full width at half maximum of the maxima is 10~MHz. In order to get an idea of the resolution, we have varied the distance of the target from 5.35 to 5.55~m in two 10~cm steps. Here, we obtain 248, 252 and 255~MHz for the respective frequencies $f_\mathrm{f}$.

From the beat frequencies, we can determine any distance of the target keeping in mind that 60~MHz corresponds to zero distance. We have varied the distance of the target between 0 and 10~m and determined its magnitude using our AFC-based FMCW LiDAR setup. Figure~\ref{fig:BeatSignal}c compares these values with the ones obtained using a commercially available device based on light-intensity modulation. The results clearly coincide with each other without any significant deviation.

\section{Discussion}\label{sec3}
The experimental results regarding the tuning linearity show that a 15~V (20~V) maximum driving signal leads to a 510~MHz (690~MHz) maximum frequency shift. If we use Eq. (\ref{eq:FrequencyShift}) with $\nu=192$~THz, $n=2.13$ \cite{Gayer.2008}, $r=25$~pm/V \cite{Mendez99}, $\eta=0.95$ \cite{MinetElectro-Optic} and $d=300$~\textmu m, the experimentally determined values nicely fit to the expected ones. We observed that the linearity of the frequency chirps decreases with decreasing rise time of the voltage signal. The largest deviation from linearity is observed in the first 10 to 20~ns of the chirps. This is independent of the rise time. We associate this with the algorithm for the determination of the frequency shifts. Before the voltage increases, there is no frequency shift. However, the algorithm requires oscillations before and after the time at which the frequency shift is to be determined. Thus, at times close to the onset of the voltage increase, the algorithm fails. Furthermore, the uncertainty of the determined frequency shift is larger as the rise time decreases. Also this can be attributed to the evaluation algorithm since it requires a certain number of oscillations to reliably determine a frequency. Most importantly, we observe that the nonlinearity of the driving voltage signal increases with decreasing rise time. This is due to the limited bandwidth (20~MHz) of the arbitrary function generator used here.
Additionally, it has been demonstrated in previous research that the quadratic electropotic effect does not require consideration up to a field strength of 65 kV/mm \cite{Luennemann03}. 
Thus, we conclude that all deviations from linearity of the frequency chirps in our experiment stem from limitations of the evaluation algorithm or from the equipment used. In other words: As long as the voltage supply delivers a linearly increasing driving signal, the adiabatic frequency converter generates a perfect linear frequency chirp. 

\begin{figure*}[h]%
\centering
\includegraphics[width = \textwidth]{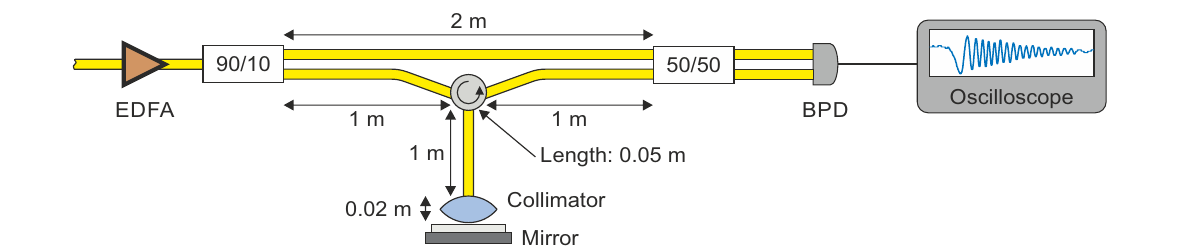}
\caption{\textbf{Schematic of the setup for the zero distance measurement }
The additional 1.1-m-long glass fiber in one of the fiber-based interferometer arms ensures a non-zero beat signal at zero distance.
To measure the frequency at zero distance, we placed a mirror directly in front of the collimator, resulting in a measured beat signal of  $f_\mathrm{B}=60$~MHz at $t_\mathrm{chirp}=130$~ns rise time. This corresponds to $\Delta t=11$~ns time delay.  
\label{fig:Distance}}
\end{figure*}

This statement is further supported by the successful application of the frequency chirps for distance measurement using FMCW LiDAR. First, the measured temporal behavior of the beat frequency, linear increase followed by a constant value and a subsequent linear decrease, displayed in Fig.~\ref{fig:Distance}a, nicely agrees with our expectation (Fig.~\ref{fig:Setup_Part2}b) under the assumption of a linear frequency chirp. 

At zero distance, we measure the beat frequency $f_\mathrm{B}=60$~MHz at $t_\mathrm{chirp}=130$~ns rise time. Considering $B=690$~MHz, this corresponds to $\Delta t=11$~ns time delay. This time delay is due to the fact that one of the fiber-based interferometer arms has an additional 1.1~m long glass fiber (refractive index 1.5). We did this on purpose in order to have a nonzero beat signal at zero distance (see also Fig.~\ref{fig:Distance}). The distance resolution can be estimated as $c_0/(2B)\approx 22$~cm \cite{BoserLidar}. Comparing the peaks of the Fourier transformation for 535 and 555~cm distances indicate that our FMCW LiDAR system seems to work bandwidth limited, i.e. the frequency chirps are indeed linear. The measurement time is limited to the decay time of the resonator, i.e. approximately to 100~ns. Thus, the maximum total path difference is 30~m which corresponds to a 15-m target distance in air. We have successfully measured distances up to 10~m. Here, we have been restricted by the size of our laboratory.
Summing up the results, we can conclude, that electro-optically driven adiabatic frequency converters can generate frequency chirps of excellent linearity while the coherence length exceeds 10~m. Furthermore, these converters are suitable light sources for real-world applications.

At first glance, the bandwidth of the chirp might not seem very impressive. However, our work showcases the huge potential of electro-optically driven adiabatic frequency converters for ultrafast and practical continuous tuning of laser light. 
We have demonstrated 690~MHz tuning in 65~ns, i.e approximately 10~GHz/\textmu s tuning rate. 
The values of our systems enabel us to compete with the fastest tunable self-injection locking laser systems \cite{KippenbergUltrafast}. However, in our study, the deviation from perfect linearity is considerably lower and it was fully limited by the equipment used and not by the tuning process itself. 
Furthermore, the huge potential regarding the tuning rate becomes even more clear if we look onto the recently realized on-chip version of an electro-optically driven adiabatic frequency converter \cite{CardenasAdiabatic}. Here 14~GHz tuning were achieved with a 14-ps-long voltage step. Such a tuning rate outperforms the values achieved with all other frequency tuning mechanisms by orders of magnitude, even the improved version of self-injection locking \cite{VahalaPockelsLaser}, and this is by far not yet the limit: It was demonstrated that lithium niobate can withstand 65~kV/mm \cite{Luennemann03} electric fields without any sign of nonlinearity in the electro-optic response. This could enable to tune the frequency of laser light by close to one THz. 
Thus, it seems to be feasible to achieve linear several-hundred-GHz-wide frequency tuning in far less than a nanosecond with tens of meters coherence length using a lithium niobate-based photonic integrated circuit. 

Of course, one has to keep in mind that tuning is possible only, if the voltage is varied on a time scale much smaller than the decay time of the light in the resonator, i.e. rise times beyond the microsecond range are difficult to achieve with lithium niobate \cite{Leidinger15}. 
Furthermore, any measurement based on adiabatic tuning of the frequency of laser light has to be performed during the decay time of the cavity and the output power is not constant.

Despite of this drawback, our results indicate that adiabatic frequency conversion is a scheme for continuously varying the frequency of laser light by hundreds of GHz in sub-nanoseconds with intrinsic tuning linearity and coherence lengths beyond 10~m. 
So far, no other method for frequency tuning achieved this. Furthermore, we demonstrated that adiabatic frequency converters are even suitable for applications such as FMCW LiDAR which sets high demands regarding the light source. In particular applications that require ultrafast frequency tuning will benefit from this tuning scheme. 
Thus, AFC can be used as a very promising platform for demanding applications in trace gas measurements, that require very high acquisition rates and tuning in to different spectral domains \cite{LongFrequency-agile}.

Although our work concentrates on the investigation of continuous adiabatic frequency conversion for generating rapid frequency chirps, our results indicate something more general: The approach of using electro-optically driven frequency converters provides some major advantages compared with tuning the laser itself. There is no competition between lasing and conversion process, i.e. the first does not limit the latter, in particular in terms of tuning rate. Applying the linear electro-optic effect provides electrically controlled rapid, linear and continuous tuning while maintaining the coherence length of the laser light. 
Further techniques following this scheme might step out of the shadow of the traditional ones. For example, single-sideband modulation. In its standard configuration, electro-optic modulators are incorporated in two coupled interferometers \cite{Fabbri13}. Modulation with an rf signal will shift the frequency of the laser light by the modulation frequency. Unwanted sidebands as well as the carrier frequency are suppressed by a careful choice of three bias voltages. The beauty of this scheme is the perfect linearity. The modulation frequency is 1:1 transferred to the frequency shift. Incorporating such a phase shifter into a sophisticated fiber loop enables 200 GHz tuning range in 5~\textmu s, i.e. 40~GHz/\textmu s tuning rate \cite{Lu18}. Here, the coherence length was estimated to exceed 1~km. Although such values seem to be very promising for large-distance measurements, compared with the abovementioned laser-based approaches, this one is far more sophisticated. 
However, recent advances in the realization of chip-integrated electro-optic frequency shifters up to 100~GHz indicate that this scheme might be simplified and ready for out-of-the-lab applications in future \cite{LoncarOn-ChipElectro-Optic}.

\section{Materials and Methods}\label{sec4}

\subsection{Device fabrication}

%\textbf{Text from Yannicks paper: Rewrite and change numbers}
%\noindent\textcolor{blue}{Wie haben wir den WGR hergestellt? Text von Yannicks paper  }\\
%
The resonators for this experiment were fabricated from a 300-$\mu$m-thick wafer of z-cut 5-$\%$-MgO-doped congruent lithium niobate.
A 150-nm-layer of chromium was applied to both sides of this wafer.
We then manufactured the resonator blank by cutting a thin cylinder with 3 mm diameter using a frequency-doubled 388-nm femtosecond pulsed laser with a 2 kHz repetition rate and an average output power of 300 mW.
Subsequently, the round blanks were precisely soldered onto a brass post for further processing.
After the soldering process we used the same femtosecond pulsed laser in combination with a computer-controlled lathe to shape the resonator. 
After the shaping process, the resonator possesses a major radius of $R = 1.2$ mm, a minor radius $r = 380$ \textmu m, and a thickness of 300 \textmu m.
Taking into account the targeted geometry, the resonator exhibits a free spectral range (FSR) of 18.6~GHz for extraordinary polarization and 18.0 GHz for ordinary polarization at a wavelength ($\lambda_L$) of 1560~nm. 
In order to achieve a high-quality surface finish and to mitigate the detrimental effects of scattering losses, an approach involving manual polishing using diamond pastes with granular dimensions as fine as 50~nm was undertaken.
The outcome was a WGR with an intrinsic quality factor of $2\times10^8$ at a wavelength of 1560~nm for e-polarized light.

\subsection{Linearity characterisation and coherent distance measurement experiment}

%\noindent\textcolor{blue}{Wie haben wir die Distanz gemessen? Setup Beschreibung  }\\

The schematics that showcases the experimental arrangement for assessing tuning linearity and the coherent ranging experiment are depicted in Fig.~\ref{fig:Setup_Part1}, respectively Fig.~\ref{fig:Setup_Part2}.
The light from a laser diode at 1.56~\textmu m wavelength is coupled into a fiber using a fiber polarization controller to get extraordinary polarized light.
Light is focused through a gradient index lens and directed into the resonator using a rutile prism (Prism 1), a process known as evanescent coupling.
We control the coupling strength at both prisms by manipulating the separation distance between each prism and the resonator via piezoelectric actuators.
In order to maintain a consistently unchanging coupling, we employ a resonator holder equipped with a temperature stabilization mechanism that operates with mK precision.
For the tuning of the cavity resonance via the Pockels effect by an applied electric field, chromium-coated top and bottom electrodes are connected to an arbitrary function generator (AFG) capable of a 20 Vpp maximum output voltage.
The frequency shifted light is extracted via the prisms and in the first prism (Prism 1) superimposed with the non-coupled pump light, which reflects off the prism base.
In the second prism (Prism 2) there is no reflected pump light and just the light from the resonator is coupled out. 
In both cases, we couple the light back into a fiber via a gradient index lens. 
The light from Prism 1 is then directed onto a photo diode connected to an oscilloscope.
With this light we did the linearity characterisation. 

For the coherent ranging experiment the light from the second prism (Prism 2) was used.
The signal is then amplified by an erbium-doped fiber amplifier (EDFA) from 10 $\mu$W up to 14 mW. 
The amplified light is then split into two paths. 
The 10 \% path serves as the reference arm and the 90 \% arm is routed through the target path.
The signal in the target path is than directed via a fiber circulator to a collimator with a focal length of $f = 7.5$ mm and a numerical aperture $NA = 0.3$, which results in a lens aperture of 4.5 mm.
From the collimator, the light is directed onto the target (a non-polished metal
plate) and reflected. 
The reflected light is focused again by the collimator in to the fiber-optical circulator.
The light from the target and the light from the reference arm are then superimposed and detected with a balanced photodetector. 
The target distance was manually adjusted to previously measured distances using a  commercially available device based on lightintensity modulation.
The beat signal is displayed and recorded on an osciloscope.

\backmatter

%\bmhead{Supplementary information}

%If your article has accompanying supplementary file/s please state so here. 

%Authors reporting data from electrophoretic gels and blots should supply the full unprocessed scans for key as part of their Supplementary information. This may be requested by the editorial team/s if it is missing.

%Please refer to Journal-level guidance for any specific requirements.

\bmhead{Acknowledgments}
We thank Karsten Buse for fruitful discussions on the manuscript.
This work was financially supported by the German Research Foundation, DFG (Grant No. BR 4194/12-1) and by the Federal Ministry of Education and Research, BMBF (Grant No. 13N16555).

\bmhead{Author contributions}
A.M. and J.O. performed the experiments and analyzed the data together with I.B. All authors discussed the results and wrote the manuscript. I.B. supervised the project.

\bmhead{Data availability}
All data that support the findings in this study are available from the
corresponding author upon reasonable request.

\bmhead{Conflict of interest}
The authors declare no competing interests.

\bibliography{sn-article}% common bib file

\begin{thebibliography}{10}
\expandafter\ifx\csname url\endcsname\relax
  \def\url#1{\burl{#1}}\fi
\expandafter\ifx\csname urlprefix\endcsname\relax\def\urlprefix{URL }\fi
\providecommand{\bibinfo}[2]{#2}
\providecommand{\eprint}[2][]{\url{#2}}
\providecommand{\doi}[1]{\url{https://doi.org/#1}}
\bibcommenthead

\bibitem{WilkeStabilized}
\bibinfo{author}{Kwee, P.} \emph{et~al.}
\newblock \bibinfo{title}{Stabilized high-power laser system for the
  gravitational wave detector advanced {LIGO}}.
\newblock \emph{\bibinfo{journal}{Opt. Express}} \textbf{\bibinfo{volume}{20}},
  \bibinfo{pages}{10617--10634} (\bibinfo{year}{2012}).

\bibitem{VahalaSearching}
\bibinfo{author}{Suh, M.-G.} \emph{et~al.}
\newblock \bibinfo{title}{Searching for exoplanets using a microresonator
  astrocomb}.
\newblock \emph{\bibinfo{journal}{Nat. Photonics}}
  \textbf{\bibinfo{volume}{13}}, \bibinfo{pages}{25--30}
  (\bibinfo{year}{2019}).

\bibitem{HidetschiAnOptical}
\bibinfo{author}{Takamoto, M.}, \bibinfo{author}{Hong, F.-L.},
  \bibinfo{author}{Higashi, R.} \& \bibinfo{author}{Katori, H.}
\newblock \bibinfo{title}{An optical lattice clock}.
\newblock \emph{\bibinfo{journal}{Nature}} \textbf{\bibinfo{volume}{435}},
  \bibinfo{pages}{321--324} (\bibinfo{year}{2005}).

\bibitem{PanAdvances}
\bibinfo{author}{Li, A.} \emph{et~al.}
\newblock \bibinfo{title}{Advances in cost-effective integrated spectrometers}.
\newblock \emph{\bibinfo{journal}{Light Sci. Appl.}}
  \textbf{\bibinfo{volume}{11}}, \bibinfo{pages}{174} (\bibinfo{year}{2022}).

\bibitem{KulakovskiiOpticalModes}
\bibinfo{author}{Bayer, M.} \emph{et~al.}
\newblock \bibinfo{title}{Optical modes in photonic molecules}.
\newblock \emph{\bibinfo{journal}{Phys. Rev. Lett.}}
  \textbf{\bibinfo{volume}{81}}, \bibinfo{pages}{2582--2585}
  (\bibinfo{year}{1998}).

\bibitem{BoserLidar}
\bibinfo{author}{Behroozpour, B.}, \bibinfo{author}{Sandborn, P. A.~M.},
  \bibinfo{author}{Wu, M.~C.} \& \bibinfo{author}{Boser, B.~E.}
\newblock \bibinfo{title}{Lidar system architectures and circuits}.
\newblock \emph{\bibinfo{journal}{IEEE Commun. Mag.}}
  \textbf{\bibinfo{volume}{55}}, \bibinfo{pages}{135--142}
  (\bibinfo{year}{2017}).

\bibitem{FujimotoTheEcosystem}
\bibinfo{author}{Swanson, E.~A.} \& \bibinfo{author}{Fujimoto, J.~G.}
\newblock \bibinfo{title}{The ecosystem that powered the translation of {OCT}
  from fundamental research to clinical and commercial impact}.
\newblock \emph{\bibinfo{journal}{Biomed. Opt. Express}}
  \textbf{\bibinfo{volume}{8}}, \bibinfo{pages}{1638--1664}
  (\bibinfo{year}{2017}).

\bibitem{Li20}
\bibinfo{author}{Li, P.} \emph{et~al.}
\newblock \bibinfo{title}{The nonlinear tuning technique of a {DFB} laser using
  frequency predistortion procedures}.
\newblock \emph{\bibinfo{journal}{Proc. SPIE}}
  \textbf{\bibinfo{volume}{11279}}, \bibinfo{pages}{1127922}
  (\bibinfo{year}{2020}).

\bibitem{NehmetallahLarge-Volume}
\bibinfo{author}{DiLazaro, T.} \& \bibinfo{author}{Nehmetallah, G.}
\newblock \bibinfo{title}{Large-volume, low-cost, high-precision {FMCW}
  tomography using stitched {DFBs}}.
\newblock \emph{\bibinfo{journal}{Opt. Express}} \textbf{\bibinfo{volume}{26}},
  \bibinfo{pages}{2891--2904} (\bibinfo{year}{2018}).

\bibitem{MingHigh-Accuracy}
\bibinfo{author}{Hariyama, T.}, \bibinfo{author}{Sandborn, P. A.~M.},
  \bibinfo{author}{Watanabe, M.} \& \bibinfo{author}{Wu, M.~C.}
\newblock \bibinfo{title}{High-accuracy range-sensing system based on {FMCW}
  using low-cost {VCSEL}}.
\newblock \emph{\bibinfo{journal}{Opt. Express}} \textbf{\bibinfo{volume}{26}},
  \bibinfo{pages}{9285--9297} (\bibinfo{year}{2018}).

\bibitem{JayaramanVCSEL}
\bibinfo{author}{John, D.~D.} \emph{et~al.}
\newblock \bibinfo{title}{Wideband electrically pumped 1050-nm {MEMS}-tunable
  {VCSEL} for ophthalmic imaging}.
\newblock \emph{\bibinfo{journal}{J. Light. Technol.}}
  \textbf{\bibinfo{volume}{33}}, \bibinfo{pages}{3461--3468}
  (\bibinfo{year}{2015}).

\bibitem{MingLaserFreq}
\bibinfo{author}{Zhang, X.}, \bibinfo{author}{Pouls, J.} \&
  \bibinfo{author}{Wu, M.~C.}
\newblock \bibinfo{title}{Laser frequency sweep linearization by iterative
  learning pre-distortion for {FMCW} {LiDAR}}.
\newblock \emph{\bibinfo{journal}{Opt. Express}} \textbf{\bibinfo{volume}{27}},
  \bibinfo{pages}{9965--9974} (\bibinfo{year}{2019}).

\bibitem{IiyamaThreee-Dimensional}
\bibinfo{author}{Ula, R.~K.}, \bibinfo{author}{Noguchi, Y.} \&
  \bibinfo{author}{Iiyama, K.}
\newblock \bibinfo{title}{Three-dimensional object profiling using highly
  accurate {FMCW} optical ranging system}.
\newblock \emph{\bibinfo{journal}{J. Light. Technol.}}
  \textbf{\bibinfo{volume}{37}}, \bibinfo{pages}{3826--3833}
  (\bibinfo{year}{2019}).

\bibitem{Jayaraman14}
\bibinfo{author}{Jayaraman, V.} \emph{et~al.}
\newblock \bibinfo{title}{Recent advances in {MEMS-VCSELs} for high performance
  structural and functional {SS-OCT} imaging}.
\newblock \emph{\bibinfo{journal}{Proc. SPIE}} \textbf{\bibinfo{volume}{8934}},
  \bibinfo{pages}{893402} (\bibinfo{year}{2014}).

\bibitem{Khan21}
\bibinfo{author}{Khan, M.~S.} \emph{et~al.}
\newblock \bibinfo{title}{{MEMS-VCSEL} as a tunable light source for {OCT}
  imaging of long working distance}.
\newblock \emph{\bibinfo{journal}{J. Optical Microsystems}}
  \textbf{\bibinfo{volume}{1}}, \bibinfo{pages}{034503} (\bibinfo{year}{2021}).

\bibitem{DrexlerAkinetic}
\bibinfo{author}{Bonesi, M.} \emph{et~al.}
\newblock \bibinfo{title}{Akinetic all-semiconductor programmable swept-source
  at 1550 nm and 1310 nm with centimeters coherence length}.
\newblock \emph{\bibinfo{journal}{Opt. Express}} \textbf{\bibinfo{volume}{22}},
  \bibinfo{pages}{2632--2655} (\bibinfo{year}{2014}).

\bibitem{LeitgebAkinetic}
\bibinfo{author}{Chen, Z.} \emph{et~al.}
\newblock \bibinfo{title}{Phase-stable swept source {OCT} angiography in human
  skin using an akinetic source}.
\newblock \emph{\bibinfo{journal}{Biomed. Opt. Express}}
  \textbf{\bibinfo{volume}{7}}, \bibinfo{pages}{3032--3048}
  (\bibinfo{year}{2016}).

\bibitem{KippenbergUltrafast}
\bibinfo{author}{Snigirev, V.} \emph{et~al.}
\newblock \bibinfo{title}{Ultrafast tunable lasers using lithium niobate
  integrated photonics}.
\newblock \emph{\bibinfo{journal}{Nature}} \textbf{\bibinfo{volume}{615}},
  \bibinfo{pages}{411--417} (\bibinfo{year}{2023}).

\bibitem{KippenbergHighDensity}
\bibinfo{author}{Li, Z.} \emph{et~al.}
\newblock \bibinfo{title}{High density lithium niobate photonic integrated
  circuits}.
\newblock \emph{\bibinfo{journal}{Nat. Commun.}} \textbf{\bibinfo{volume}{14}},
  \bibinfo{pages}{4856} (\bibinfo{year}{2023}).

\bibitem{VahalaPockelsLaser}
\bibinfo{author}{Li, M.} \emph{et~al.}
\newblock \bibinfo{title}{Integrated pockels laser}.
\newblock \emph{\bibinfo{journal}{Nat. Commun.}} \textbf{\bibinfo{volume}{13}},
  \bibinfo{pages}{5344} (\bibinfo{year}{2022}).

\bibitem{Kondratiev17}
\bibinfo{author}{Kondratiev, N.~M.} \emph{et~al.}
\newblock \bibinfo{title}{Self-injection locking of a laser diode to a high-{Q}
  {WGM} microresonator}.
\newblock \emph{\bibinfo{journal}{Opt. Express}} \textbf{\bibinfo{volume}{25}},
  \bibinfo{pages}{28167} (\bibinfo{year}{2017}).

\bibitem{SatoshiWavelength}
\bibinfo{author}{Notomi, M.} \& \bibinfo{author}{Mitsugi, S.}
\newblock \bibinfo{title}{Wavelength conversion via dynamic refractive index
  tuning of a cavity}.
\newblock \emph{\bibinfo{journal}{Phys. Rev. A}} \textbf{\bibinfo{volume}{73}},
  \bibinfo{pages}{051803} (\bibinfo{year}{2006}).

\bibitem{LipsonChanging}
\bibinfo{author}{Preble, S.~F.}, \bibinfo{author}{Xu, Q.} \&
  \bibinfo{author}{Lipson, M.}
\newblock \bibinfo{title}{Changing the colour of light in a silicon resonator}.
\newblock \emph{\bibinfo{journal}{Nat. Photonics}}
  \textbf{\bibinfo{volume}{1}}, \bibinfo{pages}{293--296}
  (\bibinfo{year}{2007}).

\bibitem{BreunigPockels}
\bibinfo{author}{Minet, Y.} \emph{et~al.}
\newblock \bibinfo{title}{Pockels-effect-based adiabatic frequency conversion
  in ultrahigh-{Q} microresonators}.
\newblock \emph{\bibinfo{journal}{Opt. Express}} \textbf{\bibinfo{volume}{28}},
  \bibinfo{pages}{2939--2947} (\bibinfo{year}{2020}).

\bibitem{Breunig22}
\bibinfo{author}{Breunig, I.} \emph{et~al.}
\newblock \bibinfo{title}{Adiabatic frequency conversion in microresonators for
  multi-wavelength holography}.
\newblock \emph{\bibinfo{journal}{Proc. SPIE}}
  \textbf{\bibinfo{volume}{11987}}, \bibinfo{pages}{1198707}
  (\bibinfo{year}{2022}).

\bibitem{CardenasAdiabatic}
\bibinfo{author}{He, X.} \emph{et~al.}
\newblock \bibinfo{title}{Electrically induced adiabatic frequency conversion
  in an integrated lithium niobate ring resonator}.
\newblock \emph{\bibinfo{journal}{Opt. Lett.}} \textbf{\bibinfo{volume}{47}},
  \bibinfo{pages}{5849--5852} (\bibinfo{year}{2022}).

\bibitem{MinetElectro-Optic}
\bibinfo{author}{Minet, Y.}, \bibinfo{author}{Zappe, H.},
  \bibinfo{author}{Breunig, I.} \& \bibinfo{author}{Buse, K.}
\newblock \bibinfo{title}{Electro-optic control of lithium niobate bulk
  whispering gallery resonators: Analysis of the distribution of externally
  applied electric fields}.
\newblock \emph{\bibinfo{journal}{Crystals}} \textbf{\bibinfo{volume}{11}}
  (\bibinfo{year}{2021}).

\bibitem{Gayer.2008}
\bibinfo{author}{Gayer, O.}, \bibinfo{author}{Sacks, Z.},
  \bibinfo{author}{Galun, E.} \& \bibinfo{author}{Arie, A.}
\newblock \bibinfo{title}{Temperature and wavelength dependent refractive index
  equations for {MgO}-doped congruent and stoichiometric
  {LiNbO\textsubscript{3}}}.
\newblock \emph{\bibinfo{journal}{Appl. Phys. B}}
  \textbf{\bibinfo{volume}{91}}, \bibinfo{pages}{343--348}
  (\bibinfo{year}{2008}).

\bibitem{Mendez99}
\bibinfo{author}{Méndez, A.}, \bibinfo{author}{García-Cabañes, A.},
  \bibinfo{author}{Diéguez, E.} \& \bibinfo{author}{Cabrera, J.~M.}
\newblock \bibinfo{title}{Wavelength dependence of electro-optic coefficients
  in congruent and quasi-stoichiometric {LiNbO\textsubscript{3}}}.
\newblock \emph{\bibinfo{journal}{Electron. Lett.}}
  \textbf{\bibinfo{volume}{35}}, \bibinfo{pages}{498} (\bibinfo{year}{1999}).

\bibitem{Luennemann03}
\bibinfo{author}{Luennemann, M.}, \bibinfo{author}{Hartwig, U.},
  \bibinfo{author}{Panotopoulos, G.} \& \bibinfo{author}{Buse, K.}
\newblock \bibinfo{title}{Electrooptic properties of lithium niobate crystals
  for extremely high external electric fields}.
\newblock \emph{\bibinfo{journal}{Appl. Phys. B}}
  \textbf{\bibinfo{volume}{76}}, \bibinfo{pages}{403} (\bibinfo{year}{2003}).

\bibitem{Leidinger15}
\bibinfo{author}{Leidinger, M.} \emph{et~al.}
\newblock \bibinfo{title}{Comparative study on three highly sensitive
  absorption measurement techniques characterizing lithium niobate over its
  entire transparent spectral range}.
\newblock \emph{\bibinfo{journal}{Opt. Express}} \textbf{\bibinfo{volume}{23}},
  \bibinfo{pages}{21690} (\bibinfo{year}{2015}).

\bibitem{LongFrequency-agile}
\bibinfo{author}{Truong, G.-W.} \emph{et~al.}
\newblock \bibinfo{title}{Frequency-agile, rapid scanning spectroscopy}.
\newblock \emph{\bibinfo{journal}{Nat. Photonics}}
  \textbf{\bibinfo{volume}{7}}, \bibinfo{pages}{532--534}
  (\bibinfo{year}{2013}).

\bibitem{Fabbri13}
\bibinfo{author}{Fabbri, S.}, \bibinfo{author}{O’Riordan, C.},
  \bibinfo{author}{Sygletos, S.} \& \bibinfo{author}{Ellis, A.}
\newblock \bibinfo{title}{Active stabilisation of single drive dual-parallel
  {Mach-Zehnder} modulator for single sideband signal generation}.
\newblock \emph{\bibinfo{journal}{Electron. Lett.}}
  \textbf{\bibinfo{volume}{49}}, \bibinfo{pages}{135} (\bibinfo{year}{2013}).

\bibitem{Lu18}
\bibinfo{author}{Lu, Z.} \emph{et~al.}
\newblock \bibinfo{title}{Broadband linearly chirped light source with narrow
  linewidth based on external modulation}.
\newblock \emph{\bibinfo{journal}{Opt. Lett.}} \textbf{\bibinfo{volume}{43}},
  \bibinfo{pages}{4144} (\bibinfo{year}{2018}).

\bibitem{LoncarOn-ChipElectro-Optic}
\bibinfo{author}{Hu, Y.} \emph{et~al.}
\newblock \bibinfo{title}{On-chip electro-optic frequency shifters and beam
  splitters}.
\newblock \emph{\bibinfo{journal}{Nature}} \textbf{\bibinfo{volume}{599}},
  \bibinfo{pages}{587--593} (\bibinfo{year}{2021}).

\end{thebibliography}

\end{document}